\documentclass[amsmath,amssymb,twocolumn,showpacs,superscriptaddress,prl,longbibliography]{revtex4-1}

\usepackage{dcolumn}
\usepackage{bm}
\usepackage{graphicx}
\usepackage{epstopdf}
\usepackage{wrapfig}
\usepackage{hyperref}
\usepackage{array} 
\usepackage{listings}
\usepackage[para,online,flushleft]{threeparttablex}
\usepackage{booktabs,dcolumn}
\usepackage{filecontents}
\usepackage{breqn}
\usepackage{xcolor}
\usepackage{siunitx}
\usepackage{bibunits}
\usepackage{lipsum}
\usepackage{upgreek} 
\usepackage{etoolbox}
\usepackage{comment}
\usepackage{soul}
\usepackage{braket}

\usepackage{orcidlink}

\makeatletter
\newcommand*{\newbibstartnumber}[1]{%
  \apptocmd{\thebibliography}{%
    \global\c@NAT@ctr #1\relax
    \addtocounter{NAT@ctr}{-1}%
  }{}{}%
}
\makeatother

\makeatletter
\newcommand\footnoteref[1]{\protected@xdef\@thefnmark{\ref{#1}}\@footnotemark}
\makeatother

\begin{document}

\preprint{APS/123-QED}

\title{Observation of the distribution of nuclear magnetization in a molecule}

\author{S.~G.~Wilkins$^{1,*,\dagger}$\orcidlink{0000-0001-8897-7227},
        S.~M.~Udrescu$^{1,*,\ddagger}$\orcidlink{0000-0002-1989-576X},
        M.~Athanasakis-Kaklamanakis$^{2,3}$,
        R.~F.~Garcia Ruiz$^{1,\S}$\orcidlink{0000-0002-2926-5569},
        M.~Au$^{4,5}$, \\
        I.~Belo\u{s}evi\'{c}$^{6}$,
        R.~Berger$^{7}$,
        M.~L.~Bissell$^{8}$,
        A.~A.~Breier$^{9}$,
        A.~J.~Brinson$^{1}$,
        K.~Chrysalidis$^{4}$,
        T.~E.~Cocolios$^{3}$, \\
        R.~P.~de~Groote$^{3}$,
        A.~Dorne$^{3}$,
        K.~T.~Flanagan$^{8,10}$,
        S.~Franchoo$^{11}$,
        K.~Gaul$^{7}$,
        S. Geldhof$^{3}$,
        T.~F.~Giesen$^{9}$, \\
        D.~Hanstorp$^{12}$
        R.~Heinke$^{4}$,
        T.~Isaev$^{13}$,
        \'{A}.~Koszor\'{u}s$^{2}$,
        S.~Kujanpää$^{14}$,
        L.~Lalanne$^{3}$,
        G.~Neyens$^{3}$,
        M.~Nichols$^{12}$, \\
        H.~A.~Perrett$^{8}$,
        J.~R.~Reilly$^{8}$,
        L.~V.~Skripnikov$^{13}$\orcidlink{0000-0002-2062-684X},
        S. Rothe$^{4}$,
        B.~van~den~Borne$^{3}$, 
        Q.~Wang$^{15}$, \\
        J.~Wessolek$^{8}$, 
        X.~F.~Yang$^{16}$,
        C.~Z\"{u}lch$^{7}$ \\~\\
\small \textit{$^{1}$Department of Physics, Massachusetts Institute of Technology, Cambridge, MA 02139, USA}\\
\small \textit{$^{2}$Experimental Physics Department, ~CERN, ~CH-1211 ~Geneva ~23, ~Switzerland}\\
\small \textit{$^{3}$KU Leuven, ~Instituut ~voor ~Kern- en ~Stralingsfysica, ~B-3001 ~Leuven, ~Belgium}\\
\small \textit{$^{4}$Systems Department, ~CERN, ~CH-1211 ~Geneva ~23, ~Switzerland}\\
\small \textit{$^{5}$Department Chemie, ~Johannes ~Gutenberg-Universitat ~Mainz, ~D-55099 ~Mainz, ~Germany}\\
\small \textit{$^{6}$TRIUMF, ~4004 ~Wesbrook ~Mall, ~Vancouver, ~BC  ~V6T ~2A3, ~Canada}\\
\small \textit{$^{7}$Fachbereich Chemie, Philipps-Universit{\"a}t~Marburg, Hans-Meerwein-Stra{\ss}e~4,~35032~Marburg,~Germany}\\
\small \textit{$^{8}$School of Physics and Astronomy, ~The ~University ~of ~Manchester, ~Manchester ~M13 ~9PL, ~United ~Kingdom}\\
\small \textit{$^{9}$Laboratory for Astrophysics, ~Institute ~of ~Physics, ~University ~of ~Kassel, ~34132 ~Kassel, ~Germany}\\
\small \textit{$^{10}$Photon Science Institute, ~The ~University ~of ~Manchester, ~Manchester ~M13 ~9PY, ~United ~Kingdom}\\
\small \textit{$^{11}$Laboratoire Irène Joliot-Curie, F-91405 ~Orsay, France}\\
\small \textit{$^{12}$Department of Physics, ~University ~of ~Gothenburg, ~SE-412 96 ~Gothenburg, ~Sweden}\\
\small \textit{$^{13}$Affiliated with an institute covered by a cooperation agreement with CERN}\\
\small \textit{$^{14}$Department of Physics, ~University ~of ~Jyväskylä, ~Survontie ~9, ~Jyväskylä, ~FI-40014, ~Finland}\\
\small \textit{$^{15}$School of Nuclear Science and Technology, ~Lanzhou ~University, ~Lanzhou ~730000, ~People's ~Republic ~of ~China}\\
\small \textit{$^{16}$School of Physics and State Key Laboratory of Nuclear Physics and Technology, ~Peking ~University, ~Beijing ~100971, ~China}\\~\\
\small \textit{Correspondence:} $^\dagger$\href{mailto:wilkinss@mit.edu}{wilkinss@mit.edu}, $^\ddagger$\href{mailto:sudrescu@mit.edu}{sudrescu@mit.edu}, $^\S$\href{mailto:rgarciar@mit.edu}{rgarciar@mit.edu}\\
\normalsize $^{*}$These authors contributed equally to this work.\\\vspace{5pt}}

\begin{abstract}

\bf Rapid progress in the experimental control and interrogation of molecules, combined with developments in precise calculations of their structure, are enabling new opportunities in the investigation of nuclear and particle physics phenomena. Molecules containing heavy, octupole-deformed nuclei such as radium are of particular interest for such studies, offering an enhanced sensitivity to the properties of fundamental particles and interactions. Here, we report precision laser spectroscopy measurements and theoretical calculations of the structure of the radioactive radium monofluoride molecule, $^{225}$Ra$^{19}$F. Our results allow fine details of the short-range electron-nucleus interaction to be revealed, indicating the high sensitivity of this molecule to the distribution of magnetization, currently a poorly constrained nuclear property,  within the radium nucleus. These results provide a direct and stringent test of the description of the electronic wavefunction inside the nuclear volume, highlighting the suitability of these molecules to investigate subatomic phenomena.
 
\end{abstract}

\maketitle

\section{Introduction}
Recent developments in the synthesis and manipulation of molecular systems are opening up a diverse range of opportunities in fundamental physics research \cite{shuman2010laser,truppe2017molecules,safronova2018search,andreev2018improved,roussy2023improved,arrowsmith2023opportunities}. Precision measurements in molecules \cite{safronova2018search,andreev2018improved,altuntacs2018demonstration,roussy2023improved,arrowsmith2023opportunities}, combined with the development of \textit{ab initio} molecular theory \cite{saue2020dirac,kallay2005approximate,kallay2020mrcc,skripnikov2020nuclear,Skripnikov:18a}, offer a compelling avenue for exploring various aspects of nuclear and particle physics \cite{andreev2018improved,safronova2018search}. The structure of certain molecular states can be highly sensitive to subtle details of electron-nucleon and nucleon-nucleon interactions within the constituent nuclei of the molecule. As these effects scale rapidly with the proton number, nuclear size, nuclear spin, and nuclear deformation \cite{auerbach1996collective,flambaum2008electric,isaev2010laser,kudashov2014ab,sasmal2016relativistic,safronova2018search,flambaum2019enhanced,gaul2019systematic,arrowsmith2023opportunities} molecules containing heavy radioactive nuclei, such as radium monofluoride, RaF, are of particular interest for fundamental physics studies  \cite{isaev2010laser,kudashov2014ab,sasmal2016relativistic,garcia2020spectroscopy,udrescu2021isotope}. The radioisotope $^{225}$Ra (half-life of $14.9$ days), with 88 protons and 137 neutrons, is expected to possess a rare nuclear octupole deformation \cite{gaffney2013studies}, boosting its sensitivity to both symmetry-conserving and symmetry-violating nuclear properties by more than three orders of magnitude with respect to stable isotopes \cite{par15,flambaum2008electric,Chu19,auerbach1996collective,isaev2010laser,kudashov2014ab,sasmal2016relativistic,gaffney2013studies,garcia2020spectroscopy,udrescu2021isotope}. The former effects are critical to guide our understanding of the nuclear force and the emergence of collective nuclear phenomena, while the latter could provide answers to some of the most pressing questions in our understanding of the universe \cite{safronova2018search,Chu19}. Measurements of the breaking of parity ($\mathcal{P}$)- and time-reversal ($\mathcal{T}$) symmetries could explain the nature of dark matter, the origin of the overwhelming imbalance between matter and anti-matter in the universe, or settle the decades-long search for charge-conjugation and parity ($\mathcal{CP}$)-violation in the strong force \cite{andreev2018improved,safronova2018search,roussy2023improved,arrowsmith2023opportunities}.

The ability to unravel nuclear and particle physics phenomena from experimental
measurements on molecules is limited by the combined precision that can be achieved experimentally and theoretically. On the theoretical side, a detailed understanding of the electronic wavefunction inside of the nuclear volume is essential to reliably extract fundamental physics information from measurements \cite{auerbach1996collective,flambaum2008electric,isaev2010laser,kudashov2014ab,sasmal2016relativistic}. Therefore, determining observables that are sensitive to the electron-nucleon interaction within the nucleus, such as the molecular hyperfine structure, is critical.

Here, we report precision laser spectroscopy measurements of the hyperfine structure of the $^{225}$Ra$^{19}$F molecule. With a lifetime on the order of just days, our results represent a major milestone in precision studies of short-lived radioactive molecules. We combine these with state-of-the-art molecular structure calculations to reveal previously unknown details of the electron-nucleus interaction in this molecule. This enables a clear observation of the effect of the $^{225}$Ra nuclear magnetization distribution on the molecular energy levels. This effect has been previously observed in atoms \cite{San23} but, to our knowledge, has never been measured before in a molecule. These findings exemplify the extreme sensitivity of the RaF molecule to properties of the Ra nucleus, and provide a direct and stringent test of the description of the electronic wavefunction within the nuclear volume.

\section{Experimental Setup}
\begin{figure*}
\includegraphics[width=2\columnwidth,height=.28\textheight]{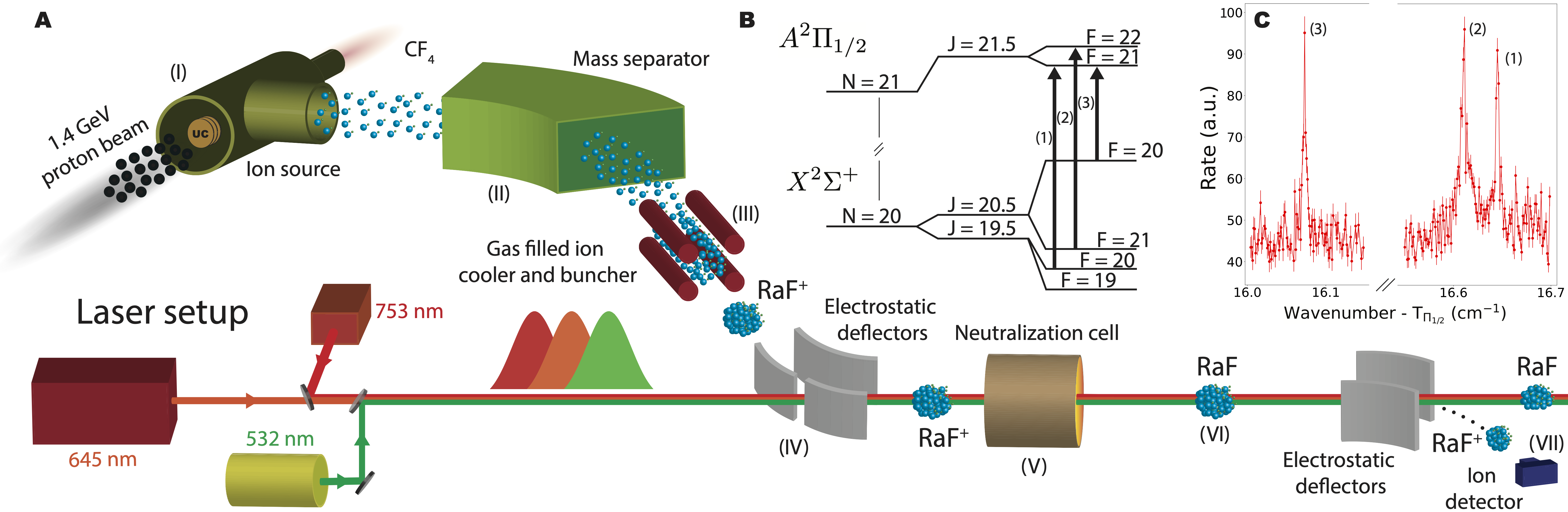}
\caption{\textbf{Experimental setup.} (\textbf{A}) Radium fluoride molecules are produced by impinging $1.4$-GeV protons on a high-temperature (T$=2000$~K) uranium carbide target, injected with CF$_4$ gas, then surfaced ionized and extracted using electrostatic fields (I). $^{225}$RaF is mass-selected (II) and trapped in a He-filled radiofrequency quadrupole (T~=~300~K) for up to $20$~ms (III). The bunched RaF ions are guided using electrostatic deflectors (IV), neutralized in a Na-filled charge-exchange cell (V), then overlapped with 3 pulsed lasers in a collinear geometry (VI). The resulting RaF ions are deflected and detected using an ion detector (VII). (\textbf{B}) Example of energy levels involved in a transition between hyperfine levels in an R-branch (not to scale). N, J and F correspond to the rotational, electronic and total angular momentum quantum numbers of the molecule (N and J are not good quantum numbers when $\mathrm{I} > 0$). Experimentally observed transitions are shown by upwards-pointing arrows and numbered. (\textbf{C}) Example of measured spectra showing the ion rate in arbitrary units (a.u.) as a function of the wavenumber of the first laser, Doppler corrected to the molecular rest frame and shifted by $T_{\Pi}$. The error bars show one standard deviation statistical uncertainty. Data points are connected by straight lines to guide the eye. The numbering on the individual peaks corresponds to the transitions shown in (\textbf{B}).} 
\label{setup}
\end{figure*}

The experiment was performed using the Collinear Resonance Ionization Spectroscopy (CRIS) setup at ISOLDE-CERN \cite{catherall2017isolde, garcia2020spectroscopy, udrescu2021isotope}. A simplified version of the experimental setup is illustrated in Fig.~\ref{setup}A. The RaF molecules were created by impinging $1.4$-GeV protons upon a uranium carbide target (UC), followed by the injection of CF$_4$ gas inside the target container at a temperature greater than $2000$~K. The RaF$^{+}$ isotopologues of interest were extracted, mass selected, then trapped and bunched in a radiofrequency quadrupole (RFQ) trap filled with He gas at room temperature for up to 20~ms. Bunches of RaF$^+$ were subsequently released, accelerated to $29908(1)$~eV, neutralized in a charge-exchange cell filled with a Na vapor at a temperature of $T\approx~500$~K and then entered the experimental interaction region. 

There, they were collinearly overlapped with three pulsed lasers in a resonance ionization scheme. The first laser employed was a Ti:Sapphire with a linewidth of $20$~MHz, which was used to excite transitions between rotational and hyperfine levels in the $X~^{2}\Sigma^{+}(v''=0)$ electronic ground state and the first excited $A~^{2}\Pi_{1/2}(v'=0)$ electronic state (Fig.~\ref{setup}B). $v''$ and $v'$ label the vibrational quantum numbers in the two electronic manifolds. Then, a pulsed dye laser (PDL), with a linewidth of $15$~GHz, was used to further excite the molecules to a higher-lying, $^2\Pi_{1/2}$, electronic state \cite{udrescu2023precision}, from which the molecules were ionized by a third, high-power $532$-nm Nd:YAG laser ($40$~mJ). The resulting RaF$^+$ ions were deflected from the neutral bunch and counted using an ion detector as a function of the first laser wavenumber, leading to the observed spectra (Fig.~\ref{setup}C). This laser ionization scheme allowed us to improve our resolution by more than two orders of magnitude and increase the signal-to-noise ratio by one order of magnitude compared to previous experiments \cite{garcia2020spectroscopy,udrescu2021isotope}, achieving a transition linewidth of $150$~MHz. This enabled an unambiguous observation of the hyperfine splitting in $^{225}$RaF due to the $^{225}$Ra nucleus (nuclear spin $\mathrm{I}=1/2$), despite its short lifetime and small rates in the interaction region (as low as 50 molecules per second in a given rotational state). A more detailed description of the experimental setup can be found in Ref.~\cite{udrescu2023precision}.

\begin{table}
\caption{\label{rotational_constants}
Rotational and hyperfine parameters in  units of cm$^{-1}$ for $X~^{2}\Sigma^{+}(v''=0)$ and $A~^{2}\Pi_{1/2}(v'=0)$ electronic manifolds of $^{225}$Ra$^{19}$F. $B''$ and $B'$ are the rotational constants for the two levels, $T_\Pi$ is the spacing between them, $p$ is the $\Lambda$-doubling parameter, while $A_\perp$ and $A_\parallel$ are hyperfine structure constants (see Materials and Methods for details). The $1\sigma$ statistical and systematic uncertainties are shown in round and square brackets, respectively. In the last two columns, values of these parameters obtained from previous experimental and theoretical studies are presented.}
\begin{ruledtabular}
\begin{tabular}{cccc}
Parameter & This work & Exp. (Lit.) & Theory (Lit.)  \\
\colrule
$B''$ & 0.192070(5)[15] & 0.19205(3)[5]\footnote{Scaled from \cite{udrescu2023precision}} & 0.1910\footnote{Scaled from \cite{zaitsevskii2022accurate}} \\ 
$A_\parallel$ & -0.5692(5)[20] & - & -0.5690[55]\footnote{Ref. \cite{skripnikov2020nuclear}} \\
$A_\perp$ & -0.5445(2)[8] & - & -0.5470[55]\textsuperscript{c}\\
\colrule
$T_\Pi$ & 13284.532(5)[20] & 13284.544(50)[20]\footnote{Ref. \cite{udrescu2021isotope,udrescu2023precision}}  & - \\
$B'$ & 0.191100(15)[45] & 0.19108(3)[5]\textsuperscript{a} & 0.1903\textsuperscript{b} \\ 
$p$ & -0.4109(15)[40] & -0.41087(9)[20]\textsuperscript{a} & -   \\
$A_\perp$ & -0.076(1)[2] & - & -0.074[1]\textsuperscript{c} \\
\end{tabular}
\end{ruledtabular}
\end{table}

\section{Results and Discussions}
The measured transitions were fitted using a rotational and hyperfine Hamiltonian for each of the two electronic states involved, using the software PGOPHER \cite{western2017pgopher}. The fit included $54$ transitions (see Fig. \ref{setup}C and Fig. \ref{spectra_fit}) and the values of the fitted rotational and hyperfine parameters obtained are shown in Table \ref{rotational_constants}. These are in excellent agreement with previous experiments, as well as with \textit{ab initio} theoretical calculations, from which they deviate by less than $1\%$ ($< 0.5$ combined standard deviation). Examples of the measured spectra together with a detailed description of the data analysis and the quantum chemistry calculations can be found in the Supplementary Materials.

\begin{figure*}[t]
\includegraphics[width=2\columnwidth,height=.4\textheight]{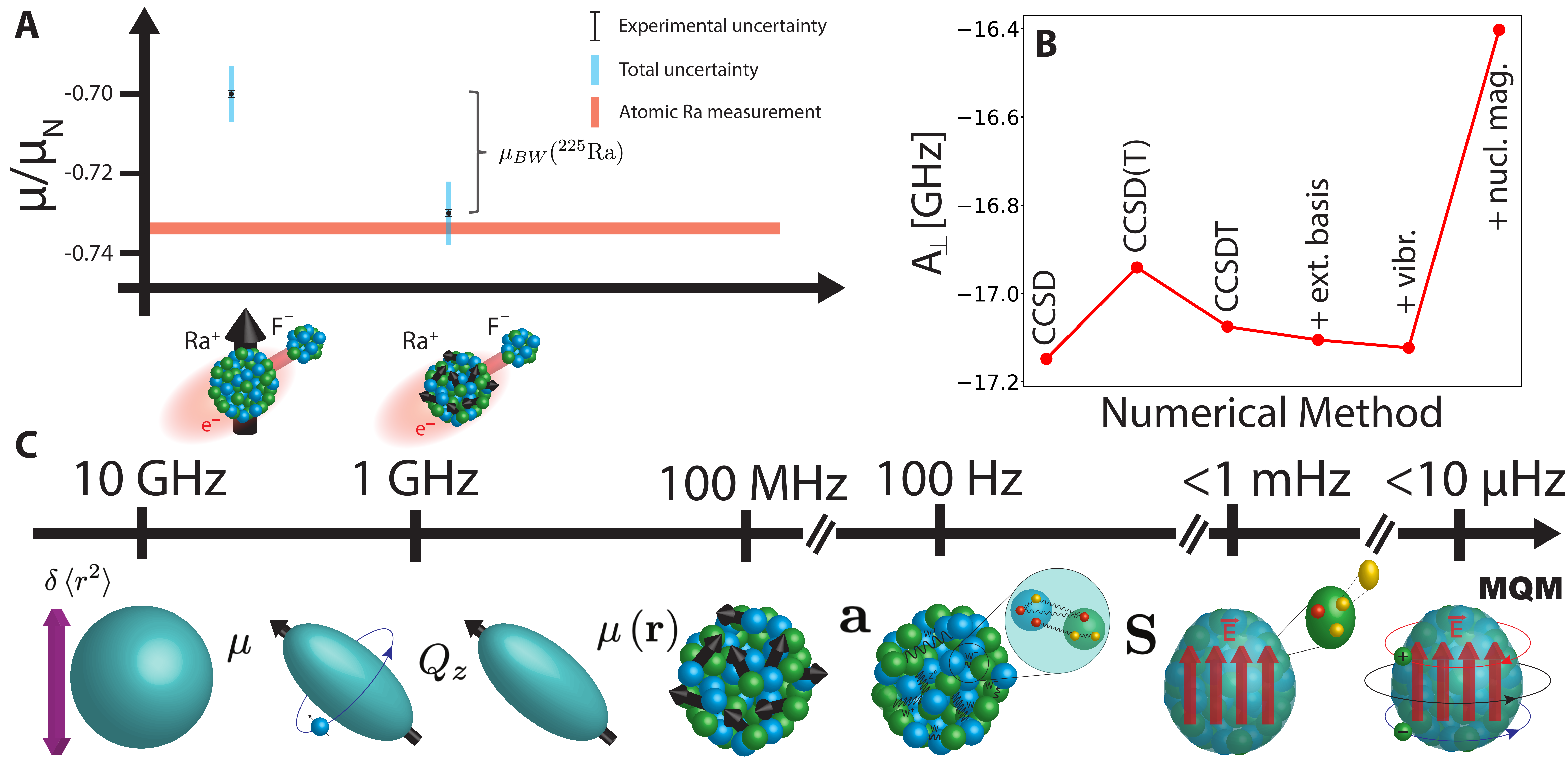}
\caption{\textbf{Nuclear effects in the RaF molecule due to the Ra nucleus.} (\textbf{A}) Extracted values of the magnetic moment of $^{225}$Ra, $\mu\left(^{225}\mathrm{Ra}\right)$, in units of nuclear magnetons, $\mu_N$, assuming the Ra nucleus is a point-like magnetic dipole (left) and accounting for the distribution of the nuclear magnetization inside of the Ra nucleus (right). The difference between the two, $\mu_{BW}$($^{225}$Ra), corresponds to the effect of the distribution of the nuclear magnetization and amounts to $\sim 5\%$ of the total value of $\mu\left(^{225}\mathrm{Ra}\right)$. The black and blue error bars are the experimental and total (experimental plus theoretical) uncertainties, respectively. The center and thickness of the orange band correspond to the previously measured value and associated uncertainty of $\mu$($^{225}$Ra) in an atom \cite{arnold1987direct}. (\textbf{B}) Evolution of the calculated $A_\perp$ for increasing levels of theoretical sophistication (see main text and the Supplementary Materials for more details) \cite{skripnikov2020nuclear}. (\textbf{C}) Order-of-magnitude estimation of nuclear effects due to Ra nucleus on the energy levels of $^{223,225}$RaF. From left to right: changes in nuclear charge radius between Ra isotopes \cite{udrescu2021isotope}, point-like magnetic dipole moment, electric quadrupole moment \cite{petrov2020energy}, distribution of nuclear magnetization, anapole moment, nuclear Schiff moment  \cite{flambaum2019enhanced,graner2016reduced}, magnetic quadrupole moment \cite{flambaum2022enhanced}. The electric and magnetic quadrupole moments are nonzero only for Ra isotopes with nuclear spin $\mathrm{I}>1/2$, such as  $^{223}$Ra.}
\label{bw_effect}
\end{figure*}

The hyperfine structure parameter of the ground state, $A_\perp$, which quantifies the strength of the coupling between the electron and the $^{225}$Ra nuclear spin, can be written as the product between the magnetic dipole moment of the $^{225}$Ra nucleus, $\mu$($^{225}$Ra), and an electronic form factor \cite{fermi1933theorie,skripnikov2020nuclear}. Using available data for the $^{225}$Ra$^+$ cation \cite{wendt1987hyperfine}, this form factor has been calculated in Ref.~\cite{skripnikov2020nuclear} for $^{225}$Ra$^{19}$F in two different ways: one in which $\mu$($^{225}$Ra) is treated as a point-like dipole moment and another one, in which the distribution of the nuclear magnetization within the $^{225}$Ra nucleus is accounted for, in a model-independent manner. Details of the extraction of the effect of the nuclear magnetization distribution and the relation between Ra$^+$ and RaF are given in the Supplementary Materials. Using these two calculated values, together with our measured $A_\perp$ parameter, the value for $\mu$($^{225}$Ra) can be precisely extracted. The obtained results are shown in Fig.~\ref{bw_effect}A, on the left, for the former case and on the right for the latter. The black error bars correspond to the experimental uncertainty, while the blue bands represent the combined theoretical and experimental uncertainty. The horizontal orange band represents the literature value of $\mu$($^{225}$Ra) and associated uncertainty, given by its thickness, obtained from an independent experiment performed on $^{225}$Ra atoms \cite{arnold1987direct}. It can be seen (Fig. 2A) that the effect of the distribution of nuclear magnetization inside of the Ra nucleus, $\mu_{BW}$($^{225}$Ra),  amounts to almost 5$\%$ of the value of $\mu$($^{225}$Ra). The $1\%$ uncertainty on the extracted value of $\mu$($^{225}$Ra) therefore corresponds to $\sim 20\%$ relative uncertainty on $\mu_{BW}$($^{225}$Ra). This level of uncertainty already has the potential to allow discrimination between simple models of the distribution of the nuclear magnetization inside the $^{225}$Ra nucleus (see Supplementary Materials). The presence of this effect, known as the Bohr-Weisskopf effect (BW) in atoms \cite{bohr1950influence}, can only be clearly observed in $^{225}$Ra$^{19}$F due to the combination of high experimental resolution and precise molecular theory (Fig. 2B).

The remarkable agreement between experiment and theory (see Table \ref{rotational_constants}), at below the $1\%$ level, reflects the reliability of the \textit{ab initio} quantum chemistry calculations, demonstrating that state-of-the-art theoretical methods are able to provide an accurate description of the electronic wavefunction within the Ra nucleus. Molecular theory is an essential ingredient for extracting fundamental physics information from precision experiments \cite{Skripnikov15b,skripnikov2016combined,safronova2018search,skripnikov2020nuclear}. Using a computational scheme similar to that used in Ref.~\cite{skripnikov2020nuclear}, the electronic parameters that provide the sensitivity of the $^{225}$Ra$^{19}$F molecule to symmetry-violating phenomena were calculated: the effective electric field $E_{\rm eff}$ acting on the electron electric dipole moment (EDM); the molecular parameter $W_{P,T}$ that characterizes the $\mathcal{P},\mathcal{T}$-violating scalar-pseudoscalar nuclear-electron interaction; the molecular constant $W_S$ that defines the interaction of the $\mathcal{P},\mathcal{T}$-violating Schiff moment of $^{225}$Ra with the electronic cloud; and the molecular parameter $W_a$ that captures the interaction between electrons and the $\mathcal{P}$-violating  nuclear anapole moment (see Supplementary Materials for details). The obtained values are shown in Table~\ref{RadiusDep} (second column). They are in good agreement with previous theoretical studies \cite{Isaev:2012,Borschevsky:13,kudashov2014ab,sasmal2016relativistic} (third column of Table~\ref{RadiusDep}), but are more precise, by as much as an order of magnitude, mainly due to a more complete treatment of correlation effects for all electrons of RaF.

The values of hyperfine structure (HFS) constants, $A_\perp$ and $A_\parallel$, as well as the symmetry-violating electronic form factors, $E_{\rm eff}$, $W_{P,T}$, $W_S$, and $W_a$, strongly depend on the electronic density behaviour in the vicinity of the $^{225}$Ra nucleus \cite{safronova2018search}. However, unlike the HFS constants measured in this work, the other computed parameters, which provide the sensitivity to symmetry-violating phenomena, cannot be measured experimentally. Hence, HFS measurements are essential for benchmarking \textit{ab initio} theoretical calculations, and critically, they allow a reliable prediction of the molecular sensitivity to symmetry-violating properties. As an illustrative example of the strong connection between the HFS constants and symmetry violating electronic form factor, it can be shown that the $A_\perp$ and $A_\parallel$ parameters can be approximately related to $E_{\rm eff}$ using a semi-empirical approximation \cite{Kozlov:97c},  $E_{\mathrm{eff}}=\alpha\sqrt{AA_d}$, where $A\equiv(A_\parallel+2A_{\perp})/3$, $A_d\equiv(A_\parallel-A_{\perp})/3$ and $\alpha \approx 0.0313$ GV/(cm MHz) is an approximate proportionality constant that can be obtained by simple numerical calculations \cite{Kozlov:97c}. This approach provides a value of $E_{\rm eff}=63.2$ GV/cm, which is within 20\% of our accurate state-of-the-art {\it ab initio} calculations, reported in Table~\ref{RadiusDep}.

\begin{table}[]
\caption{Calculated symmetry-violating electronic form factors in $^{225}$Ra$^{19}$F. The parameter names and their units are shown in the first column. The second and third columns show the values of the parameters calculated in this work and previous studies (see Supplementary Materials for details). The numbers in square brackets correspond to one standard deviation uncertainty (where available).}
\label{RadiusDep}
\begin{tabular}{lll}
 & This work & Previous work \\
\hline
\hline
$E_{\rm eff}$ (GV/cm) & -53.3[9] & \begin{tabular}{l} 
-52.8[53]\textsuperscript{a}  \\
-52.5[52]\textsuperscript{b} \\
-56.9\textsuperscript{c}  \\
-50.8\textsuperscript{d}  \\
-54.2[54]\textsuperscript{e}  \\
\end{tabular} \\
\hline
$W_{P,T}$ ($h$ kHz) & -144.3[14] & \begin{tabular}{l} 
-139[14]\textsuperscript{a} \\ 
-141.2[140]\textsuperscript{b} \\ 
-152.5\textsuperscript{c} \\  
-138\textsuperscript{d} \\  
\end{tabular} \\
\hline
$W_a$ ($h$ Hz) & 1694[17] & 
\begin{tabular}{l} 1700[170]\textsuperscript{a}  \\  
2100[315]\textsuperscript{f} \\
1641[246]\textsuperscript{g}
\end{tabular} \\
\hline
$W_S$ $(e/(4\pi\epsilon_0 a_0^4))$ & -20900[2100] & 
\begin{tabular}{l}
-22130[2213]\textsuperscript{a,h}  \\
-19148\textsuperscript{d,h}  
\end{tabular} \\
\hline
\end{tabular}
\\
\textsuperscript{a} Ref. \cite{kudashov2014ab}, \textsuperscript{b} Ref. \cite{sasmal2016relativistic}, \textsuperscript{c} Ref. \cite{Sunaga:2019}, \textsuperscript{d} Ref. \cite{Gaul:2020},\\
\textsuperscript{e} Ref. \cite{Cheng:2021}, \textsuperscript{f} Ref. \cite{Isaev:2012}, \textsuperscript{g} Ref. \cite{Borschevsky:13}, \textsuperscript{h} Ref. \cite{Flambaum:2019b}
\end{table}

Using the calculated hyperfine and $\mathcal{P},\mathcal{T}$-odd electronic form factors, the contributions of various nuclear effects to the molecular spectra of $^{223,225}$RaF are illustrated in Fig~\ref{bw_effect}C. Electroweak nuclear properties related to W- and Z-boson exchange within the Ra nucleus, such as the anapole moment, are expected to be on the order of $100$~Hz. Experiments on stable molecules are already able to achieve and exceed this level of precision \cite{safronova2018search}. The best upper limits on nuclear CP-violation effects \cite{graner2016reduced} will translate to shifts on the order of mHz in $^{225}$RaF. Hence, measurements with this precision or better in $^{225}$RaF, which is already within reach of existing atomic and molecular techniques \cite{arrowsmith2023opportunities}, can establish record bounds on hadronic CP-violation \cite{andreev2018improved,roussy2023improved}. Molecules containing Ra nuclei therefore represent some of the most compelling systems for discovering $\mathcal{CP}$-violation in the strong force \cite{safronova2018search,arrowsmith2023opportunities}.   

\section{Conclusions and outlook}
The hyperfine structure of $^{225}$Ra$^{19}$F was measured, revealing the high sensitivity of this molecule to the properties of the $^{225}$Ra nucleus. The observation of the distribution of the nuclear magnetization effect in a molecule was possible thanks to the combined precision of our experiment and high accuracy of quantum chemistry calculations, which are now reaching the sub-percent level. Improving the precision of these calculations by a factor of $\sim 2-3$ would already enable different nuclear magnetization models to be distinguished between, at below the $10 \%$ level (see Ref. \cite{skripnikov2020nuclear} and the Supplementary Materials), facilitating stringent tests of nuclear theory. We hope that our experimental results will motivate the development of higher-accuracy molecular and nuclear structure calculations. 

Our findings lay the groundwork for using these molecules in future studies of higher-order symmetry-conserving nuclear moments such as the nuclear magnetic octupole moment \cite{de2022precision}, electric hexadecapole \cite{beloy2008hyperfine,xiao2020hyperfine} or electric quadrupole shift (higher-order correction to the electric quadrupole interaction, due to electron penetration into the nucleus) \cite{Ste10}. The former has never been measured in a molecule, while the latter two have not been observed in any atom or molecule so far. All of these properties are enhanced in molecular systems containing heavy, octupole-deformed nuclei \cite{safronova2018search,arrowsmith2023opportunities}. Together with the distribution of the nuclear magnetization, they can provide valuable information about the behaviour of protons and neutrons within atomic nuclei which is important for elucidating the microscopic origin of collective nuclear phenomena. Observables that are sensitive to the neutron distribution would be key to our understanding of nuclear matter and constrain properties of neutron stars \cite{yang2022laser}. Our measurements provide critical information on the rotational and hyperfine structure of $^{225}$Ra$^{19}$F, which, complemented by the calculated electronic form factors, represent a major milestone towards future experimental developments that aim to use these molecules for fundamental physics studies \cite{safronova2018search}. 

Presently, ISOLDE can produce on the order of $10^6$ $^{225}$RaF molecules per second. The current challenge involves maximizing the number of molecules that can be cooled and prepared in a particular quantum state. Our results lay the foundation upon which these experimental developments can be built. The high sensitivity of $^{225}$RaF molecules translates to needing only a few molecules to provide competitive bounds on CP violation. A measurement with $1$ mHz precision, which would set record bounds on hadronic CP violation, can be achieved, for example, with just $100$ molecules per second, a coherence time of $100$ ms, and a total experimental integration time of $15$ days. There exist viable pathways to reach and exceed the aforementioned numbers of molecules, coherence time and integration time, which are currently being pursued by multiple research groups worldwide \cite{arrowsmith2023opportunities}.

\bibliography{references}

\section{Supplementary Materials}

\subsection{A. Data analysis}

Each peak used in the determination of the rotational and hyperfine Hamiltonian parameters was fit with a Voigt profile plus a constant background, using the LMFIT Python package. The number of peaks in a given scan was chosen based on the reduced-$\chi^{2}$ of the fit. The obtained central value of each peak, together with its associated uncertainty, were input into PGOPHER where they were fit with the effective Hamiltonians described below. The main sources of systematic uncertainty in our experiment were (expressed as an uncertainty on the wavenumber in brackets): variations in the ion beam extraction voltage ($1.2 \times 10^{-4}$ cm$^{-1}$), changes in the beam energy during the charge-exchange process ($10^{-4}$ cm$^{-1}$), uncertainties in the measurement of the Rb reference frequency by the wavemeter ($10^{-4}$ cm$^{-1}$), presence of stray magnetic and electric fields ($< 10^{-5}$ cm$^{-1}$) and AC Stark shifts due to the presence of the second- and third-step lasers ($5 \times 10^{-4}$ cm$^{-1}$). These uncertainties were added in quadrature to the statistical uncertainty for each fitted line in the spectra, before performing the PGOPHER fit.

For the $X~^{2}{}{\Sigma}^{+}$ electronic state, the employed rotational and hyperfine Hamiltonian is given by:
\begin{equation}
    H_{X~^{2}{}{\Sigma}^{+}}^{\mathrm{rot}} = \left(B''-D''\bm{N}^2\right)\bm{N}^2+\gamma \bm{N}\cdot \bm{S} + b''\mathbf{I}\cdot\mathbf{S} + c''I_zS_z,
    \label{ground_state_H} 
\end{equation}
where $B''$ is the rotational constant, $D''$ is the centrifugal distortion constant, $\gamma$ is the spin-rotation constant and $b''$ and $c''$ are hyperfine constants due to the $^{225}$Ra nucleus. These parameters can be related to $A_\perp$ and $A_\parallel$ of the $X~^{2}{}{\Sigma}^{+}$ state from \cite{skripnikov2020nuclear} using: $A_\perp = b''$ and $A_\parallel-A_\perp=c''$ \cite{denis2022benchmarking}. $\bm{N} = \bm{J} - \bm{S}$, $\bm{S}$ and $\bm{I}$ are the molecular rotational operator (excluding the electron and nuclear spin), the electron spin operator and the nuclear spin operator, respectively, while $S_z$ and $I_z$ are the z-component of the latter two.

The excited electronic state, $A~^{2}{}{\Pi}$, was described by the effective rotational and hyperfine Hamiltonians:

\begin{equation}
    \begin{split}
        H_{A~^{2}{\Pi}}^{\mathrm{rot}} & = T_\Pi + A_\Pi L_zS_z + \left(B'-D'\bm{N}^2\right)\bm{N}^2 - \\
        & - \frac{1}{2}\left\{\frac{p}{2}+p_D \bm{N}^2,N_+S_+e^{-2i\phi}+N_-S_-e^{2i\phi}\right\} + \\
        & + \frac{1}{2}d\left(e^{-2i\phi}I_+S_++e^{2i\phi}I_-S_-\right),
    \label{excited_state_H} 
    \end{split}
\end{equation}
where $\left\{O,Q\right\} = OQ+QO$, $L_z$ is the z-component of the electron orbital momentum operator in the molecular rest frame, $T_\Pi$ represents the energy difference between the origins of the $\nu' = 0$ vibrational level of the $^2\Pi$ electronic manifold and the origins of the corresponding isovibrational level of the $X~^{2}{}{\Sigma}^{+}$ electronic manifold, while $A_\Pi$ is the spin-orbit interaction. The $^2\Pi$ electronic manifold, gets split, in a Hund case (a) picture, into a $^2\Pi_{1/2}$ and a $^2\Pi_{3/2}$ electronic levels due to the spin-orbit coupling, separated by $A_\Pi$. As rovibronic transitions to the $A^2\Pi_{3/2}$ electronic state were not measured, it was not possible to constrain both $T_\Pi$ and $A_\Pi$ simultaneously, therefore $A_\Pi$ was kept fixed at its previously measured value of $2067.6$ cm$^{-1}$ \cite{garcia2020spectroscopy}. $p$ is the $\Lambda$-doubling parameter and $p_D$ is the centrifugal distortion correction to $p$. Finally, $d$ is a hyperfine structure constant due to the $^{225}$Ra nucleus. In a $A^{2}{}{\Pi}$ state, this is related to the $A_\perp$ parameter from \cite{skripnikov2020nuclear} by $A_\perp = d$ \cite{denis2022benchmarking}. $N_{\pm}$, $S_{\pm}$ and $I_{\pm}$ are the raising and lowering operators for the $\bm{N}$, $\bm{S}$ and $\bm{I}$ operators defined above and $\phi$ is the polar angle around the molecular axis, defined in the molecule's rest frame. The molecular parameters, extracted from fitting the above Hamiltonians to the data, correspond to the ground vibrational level of each electronic manifold \cite{udrescu2023precision}. The hyperfine splitting due to the fluorine nucleus ($\mathrm{I} = 1/2$) was predicted to be much below $100$ MHz and hence was not observed given our current spectroscopic resolution. Therefore, the corresponding hyperfine Hamiltonian was not included in the analysis. 

For the fitting procedure, $\gamma$, $D'$, $D''$ and $p_D$ parameters were each sampled from a Gaussian with mean and standard deviation given by the values of the corresponding parameters of the $^{226}$Ra$^{19}$F molecule \cite{udrescu2023precision}, scaled accordingly using the reduced mass of the two isotopologues. After sampling, these parameters were kept constant during the fitting procedure, and the values of the other rotational and hyperfine parameters were extracted. The sampling and subsequent fitting was repeated $1000$ times, and the obtained average value and standard deviation of the fitted parameters are reported in Table~\ref{rotational_constants}.

\begin{figure*}[t]
\includegraphics[width=2.05\columnwidth,height=.45\textheight]{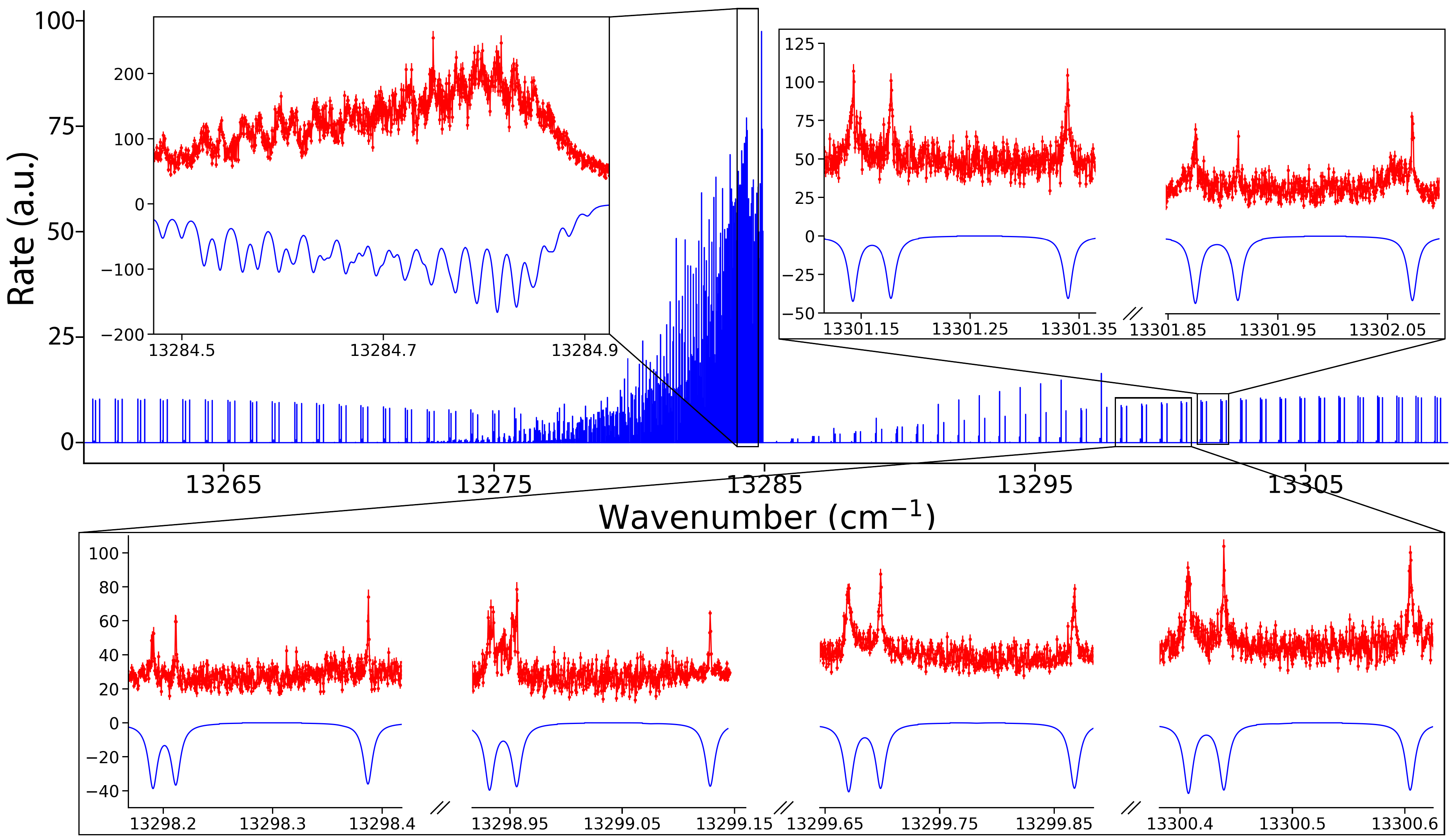}
\caption{\textbf{Example of measured spectra for the $0' \leftarrow 0''$ transitions.} In the center, in blue, the fitted combined hyperfine and rovibronic spectrum of $^{225}$RaF obtained for $J\le 100$, over a range of $\sim 50$ cm$^{-1}$ is presented. Figures in magnified views show measured spectra for different regions in frequency space. The connected red dots show the experimental data, while the continuous blue lines represents the best fits to the data. The errorbars show one standard deviation statistical uncertainty. The values on the x-axis correspond to the wavenumber of the first laser used in the resonance ionization scheme, Doppler corrected to the molecular rest frame. The rate on the y-axis is given in arbitrary units (a.u.).}
\label{spectra_fit}
\end{figure*}

Examples of the measured spectra of $^{225}$Ra$^{19}$F together with the obtained best fit are shown in Fig.~\ref{spectra_fit}, where the experimental data is shown in red, while the best fit is shown in blue. In the center, the full simulated spectrum of the transitions over a range of $\sim 50$ cm$^{-1}$ is shown. The x-axis shows the wavenumber of the first-step laser, Doppler corrected to the molecular rest frame, while the y-axis shows the rate in arbitrary units (a.u.). The splitting of each rotational line into $3$ hyperfine components can be clearly observed in the R-branch spectra around $13300$ cm$^{-1}$. 

\subsection{B. Nuclear magnetization distribution effect in Ra$^+$ and RaF}

The magnetic dipole hyperfine structure (HFS) constant can be expressed using the following parametrization~\cite{bohr1950influence}:
\begin{equation}
    A= A^{(0)} - A^{\rm BW}.
\label{HFSbw}    
\end{equation}
Here, $A^{(0)}$ represents the HFS constant in the point-like nuclear magnetic dipole moment approximation, and $A^{\rm BW}$ quantifies the contribution of the finite nuclear magnetization distribution to the HFS constant, commonly known as the Bohr-Weisskopf (BW) effect. In Ref.~\cite{skripnikov2020nuclear}, it was demonstrated that the BW correction, $A^{\rm BW}$, for heavy atoms and molecules (including systems with complex electronic structures) can be factorized as follows (see Eq.~(29) of Ref.~\cite{skripnikov2020nuclear} for details):
\begin{eqnarray}
 \label{AparBWfactorization}
 A^{\rm BW}  \approx   E B_s.
\end{eqnarray}
Here, $E$ represents a pure electronic factor, which is independent of the nuclear magnetization distribution and is solely determined by the electronic structure. All information regarding the nuclear magnetization distribution can be encoded in the electronic state-independent parameter, $B_s$. This parameter has a well-defined physical meaning~\cite{skripnikov2020nuclear} being proportional to the BW effect contribution to the $A$ constant of the hydrogen-like (H-like) ion. Therefore, in cases where measurements of H-like ions are available for a given isotope, $B_s$ can be obtained almost directly~\cite{Skripnikov:2022} due to the very high accuracy of the theoretical description of such ions.

In Ref.~\cite{skripnikov2020nuclear}, $A^{(0)}$ and $E$ were computed for the ground electronic state $7s~^2S_{1/2}$ of the $^{225}$Ra$^+$ cation. By combining these theoretical values with the experimental value~\cite{Neu:1988,wendt1987hyperfine,Dammalapati:2016} of the $A$ constant, the parameter $B_s$ was determined. The electronic $A^{(0)}$ and $E$ constants were also calculated for the excited electronic state $7p~^2P_{1/2}$ of $^{225}$Ra$^+$. Using these values, along with the extracted value of $B_s$, the BW contribution to the $A$ constant for this state was computed according to Eq.~(\ref{AparBWfactorization}). The resulting value of $A$($7p~^2P_{1/2}$) was in good agreement with the available experimental value~\cite{Neu:1988,wendt1987hyperfine,Dammalapati:2016}, with a deviation of about 0.1\%, though the theoretical uncertainty was estimated at 1\%. The BW effect contribution to $A$($7p~^2P_{1/2})$ was $1.4\%$. Interestingly, the Bohr-Weisskopf effects for the $7s~^2S_{1/2}$ and $7p~^2P_{1/2}$ states are induced by different harmonics, $s_{1/2}$ and $p_{1/2}$, respectively~\cite{skripnikov2020nuclear}. However, as explained in Ref. \cite{skripnikov2020nuclear},  the same constant $B_s$ can be utilized in both cases due to properties of the solutions to the Dirac equation and the symmetry of the magnetic dipole hyperfine interaction operator. This is particularly important for systems with complex electronic structures, where both harmonics contribute simultaneously. The validity of factorization~(\ref{AparBWfactorization}) was also numerically confirmed by considering various nuclear magnetization distribution models~\cite{skripnikov2020nuclear,Prosnyak:2021}. Finally, the semi-empirically extracted value of $B_s$ was employed to predict the BW effect in the $^{225}$RaF molecule for both the ground and first excited electronic states. The uncertainty on the deduced BW effects using this approach is limited by the uncertainty in the electronic structure calculation of $A^{(0)}$ and $E$ of Ra$^+$, estimated to be $1\%$ \cite{skripnikov2020nuclear}. Given that the BW effect amounts to about $5\%$ of the $A$ value of the $7s~^2S_{1/2}$ ground state in Ra$^+$ \cite{skripnikov2020nuclear}, the uncertainty in the value of the BW effect obtained using this semi-empirical method is $\sim 20$\%. 

It is also possible to estimate the $B_s$ parameter using simple nuclear magnetization distribution models. The simplest model is a uniformly magnetized sphere, while a more accurate model is a Woods-Saxon (WS) model. The WS model was used in Ref.~\cite{ginges2017ground} to calculate the BW effect in $^{225}$Ra$^+$. By combining the value of $B_s^{Ball}$ with the data from Ref.~\cite{ginges2017ground} and using the factorization property (\ref{AparBWfactorization}), one can obtain $B_s^{WS}$. Using Eq.~(\ref{AparBWfactorization}) and the values of the $B_s$ parameter from different nuclear magnetization distribution models, one can calculate the value of the $A_{||}^{\rm BW}$ and $A_{\perp}^{\rm BW}$ constants for the ground electronic state of the $^{225}$RaF molecule. The obtained results are given in Table \ref{RaFBW}. The uncertainty of the BW effect depends on the nuclear model used ~\cite{Senkov:02,shabaev1997ground} and it is expected that future nuclear structure calculations, employing more realistic models, will allow a prediction of this effect with quantifiable uncertainties. An improvement in the accuracy of the electronic structure calculations by only a factor of $\sim 2-3$, combined with our experimental results, would allow the study of the distribution of the nuclear magnetization with a relative precision of better than $10\%$. 

\begin{table}[]
\caption{The BW contribution to the hyperfine structure constants, $A_{||}$ and $A_{\perp}$, (in MHz) for the ground electronic state of the $^{225}$RaF molecule, using different models of the nuclear magnetization.}
\label{RaFBW}
\begin{tabular}{lll}
\hline
\hline
 Model        &  $A_{||}^{\rm BW}$    & $A_{\perp}^{\rm BW}$  \\
\hline              
Ball          & -537                  & -529  \\
WS            & -830                  & -818 \\
Semi-empirical~\cite{skripnikov2020nuclear} & -730 & -720 \\
\hline
\hline
\end{tabular}
\end{table}

\subsection{C. Computational methods}
The Hund case (c) matrix elements $\langle ^2\Sigma_{1/2}|J^{e}_{+}| ^2\Sigma_{-1/2}\rangle$ and $\langle ^2\Pi_{1/2}|J^{e}_{+}| ^2\Pi_{-1/2}\rangle$, where $J^{e}_{+}$ is the $x+iy$ component of the body-fixed total electronic angular momentum, can be related to the spin-rotational and $\Lambda$-doubling parameters in Hund cases (b) and (a) in the above Hamiltonians as follows: $\langle ^2\Sigma_{1/2}|J^{e}_{+}| ^2\Sigma_{-1/2}\rangle = 1-\frac{\gamma_e}{2B''_e}$ and $\langle ^2\Pi_{1/2}|J^{e}_{+}| ^2\Pi_{-1/2}\rangle = \frac{p_e}{2B'_e}$, where $\gamma_e$, $p_e$, $B'_e$ and $B''_e$ refer to vibrationally independent molecular parameters (see Ref. \cite{udrescu2023precision} for details). These matrix elements were calculated herein and the values obtained are reported in Table~\ref{TcalculationsJel}, exhibiting an excellent agreement with the experiment \cite{udrescu2023precision}, at the $0.1\%$ level (the values of these parameters are the same for $^{226}$RaF \cite{udrescu2023precision} and $^{225}$RaF). The following scheme was used for the calculations. First, correlation calculations were performed employing the relativistic coupled-cluster approach with single- and double-excitation amplitudes (CCSD) within the Dirac-Coulomb Hamiltonian~\cite{gomes2019dirac,saue2020dirac}. Here, 69 electrons of RaF were included in the correlation treatment and the extended uncontracted all-electron triple-zeta extAE3Z basis set (based on AE3Z~\cite{dyall2016relativistic,Dyall:2009} by Dyall) developed in Ref.~\cite{skripnikov2020nuclear} was used.
It includes $[38s\, 33p\, 24d\, 14f\, 7g\, 3h\, 2i]$ Gaussian-type functions for Ra and corresponds to the uncontracted AE3Z~\cite{dyall2016relativistic,Dyall:2009} basis set on F.  
To account for effects of larger basis sets, a correction was taken as the difference between values of the matrix elements under consideration using the extended quadruple-zeta extAE4Z~\cite{skripnikov2020nuclear} basis set and the extAE3Z basis set. These calculations were performed at the 27-electron CCSD level using the valence part of the generalized relativistic effective-core potential approach~\cite{titov1999generalized,mosyagin2010shape,mosyagin2016generalized}. 
Next, higher-order correlation effects were implemented through two contributions. The first of which was calculated as the difference between the results obtained within the CCSD and partial iterative triple-excitation amplitudes (CCSDT-3~\cite{kallay2005approximate}) model compared to the CCSD method. In this calculation, $35$ electrons of RaF were correlated using the special compact basis set for Ra constructed using the approach developed in Refs.~\cite{skripnikov2013relativistic,skripnikov2016combined,skripnikov2020nuclear} and comprising $[8s\, 8p\, 7d\, 7f\, 4g\, 2h]$ contracted Gaussian functions, while for F we used the aug-cc-pVDZ-DK~\cite{Kendall1992,de2001parallel} basis set. 
Then, correlation effects up to the CC with full iterative triple- and perturbative quadruple-excitation amplitudes CCSDT(Q)~\cite{kallay2005approximate,kallay2020mrcc} approach were implemented. For this contribution, $27$ electrons of RaF were correlated and the reduced compact basis set reduced for Ra to $[8s\, 8p\, 7d\, 4f]$ contracted Gaussian functions was used. Finally, the effect of the Gaunt interelectron interaction at the self-consistent level was computed. In all of the calculations, the equilibrium Ra--F distance was used.

\begin{table}
\caption{Contribution to the theoretically calculated $\langle ^2\Sigma_{1/2}|J^{e}_{+}| ^2\Sigma_{-1/2}\rangle$ and $\langle ^2\Pi_{1/2}|J^{e}_{+}| ^2\Pi_{-1/2}\rangle$ matrix elements. The associated experimental values extracted from \cite{udrescu2023precision} are shown in the last row. The numbers in round (square) brackets correspond to $1\sigma$ statistical (systematic) uncertainty.}
\begin{tabular}{lcc}
\hline
\hline
Contribution   & $\langle ^2\Sigma_{1/2}|J^{e}_{+}| ^2\Sigma_{-1/2}\rangle$ & $\langle ^2\Pi_{1/2}|J^{e}_{+}| ^2\Pi_{-1/2}\rangle$ \\
\hline
CCSD                           & 0.98355	& -1.05444    \\
Basis correction                   & -0.00010   & -0.00655  \\
CCSDT-3 $-$ CCSD           & 0.00050    & -0.00728  \\
CCSDT(Q) $-$ CCSDT-3       & 0.00001    & -0.00407  \\
Gaunt                              & 0.00057    & 0.00151   \\
\\
Total                              & 0.98453[67]    & -1.0708[97]  \\
Experiment                         & 0.98475(13)[34] & -1.07335(21)[45] \\
\hline
\end{tabular}
\label{TcalculationsJel}
\end{table}

The $\mathcal{P,T}$-breaking interaction induced by the $e$EDM can be described by the following Hamiltonian~\cite{MartenssonPendrill1987,Lindroth89}:
\begin{eqnarray} 
  H_d^{{\rm eff}}= d_e\sum_a  2i  c\gamma^0_a\gamma_a^5\bm{p}_a^2,
 \label{Wd2}
\end{eqnarray}
where index $a$ denotes each electron, $\bm{p}$ is the electron momentum operator, $c$ is the speed of light, $d_e$ is the electron EDM and $\gamma^0$ and $\gamma^5$  are Dirac matrices defined according to Ref.~\cite{Khriplovich91}. These matrices are related through $\gamma^5=-i\gamma_0\gamma_1\gamma_2\gamma_3$.
This interaction can be characterized by the molecular constant $W_d$:
\begin{equation}
\label{matrelem}
W_d = \frac{1}{\Omega}
\langle \Psi|\frac{H_d}{d_e}|\Psi
\rangle.
\end{equation}
%
In these designations, the effective electric field acting on the $e$EDM is $E_{\rm eff}=W_d|\Omega|$.
Another possible source of the $\mathcal{P,T}$-violation is the scalar-pseudoscalar nucleus-electron interaction given by the following Hamiltonian (see Ref.~\cite{sushkov1978parity}):
\begin{eqnarray}
  H_{\rm s}=i\frac{G_F}{\sqrt{2}}Zk_{\rm s}\sum_a \gamma^0_{a}\gamma^5_{a}\rho_N(\textbf{r}_a),
 \label{Htp}
\end{eqnarray}
where $G_{\mathrm{F}}~=~2.22249 \cdot 10^{-14}~a.u$ is the Fermi-coupling constant, $Z$ is the charge of the heavy nucleus ($Z=88$ for $^{225}$Ra in our case), $\rho_N(\textbf{r})$ is the nuclear density normalized to unity and $\mathbf{r}$ is the electron radius vector with respect to the heavy atom nucleus under consideration.
This interaction is characterized by the molecular parameter $W_{P,T}$:
\begin{equation}
\label{WTP}
W_{P,T} = \frac{1}{\Omega}
\langle \Psi|\frac{H_{\rm s}}{k_{\rm s}}|\Psi
\rangle.
\end{equation}
The electron-nucleus $\mathcal{P}$-odd interaction Hamiltonian is defined as:
\begin{equation}
 H_{P} = \kappa\frac{G_{\mathrm{F}}}{\sqrt{2}}
\bm{\alpha}\cdot\bm{I}
\rho_N(\textbf{r}).
\label{HP}
\end{equation}
The main contributions to this effect in $^{225}$RaF are the Z$^0$-boson exchange between the unpaired electron and the $^{225}$Ra nucleus and the interaction of the unpaired electron with the nuclear anapole moment. These effects are characterized by the dimensionless constant $\kappa$. By averaging this Hamiltonian over the electronic wavefunction of the molecule, the following rotational and hyperfine effective Hamiltonian is obtained \cite{Flambaum:85b}:
\begin{equation}
H_{\rm eff} =  
(W_{a}\kappa)\mathbf{n\times S}^{\prime}\cdot
\mathbf{I},
\label{HEFF}
\end{equation}
where $\mathbf{n}$ is the unit vector directed from the Ra nucleus to F and $W_{a}$ is a molecular parameter given by:
\begin{equation}
     W_{a}=\frac{G_{\mathrm{F}}}{\sqrt{2}}
     \left\langle ^2\Sigma_{1/2} \left\vert
     \rho_N(\textbf{r}) {\alpha_+}
     \right\vert ^2\Sigma_{-1/2} \right\rangle.
\label{W_a}
\end{equation} 

To calculate $E_{\rm eff}$, $W_{P,T}$ and $W_{a}$, a similar scheme to that in Ref.~\cite{skripnikov2020nuclear} was used to calculate hyperfine structure constants. First, correlation calculations were performed employing the relativistic coupled-cluster approach with single-, double- and perturbative triple-excitation amplitudes, CCSD(T), within the Dirac-Coulomb Hamiltonian~\cite{gomes2019dirac,saue2020dirac}. All 97 electrons of RaF were included in the correlation treatment using the extAE3Z basis set. The virtual energy cutoff was set to 10,000 $E_h$. The significance of the high energy cutoff for properties that depend on the behavior of the wavefunction near the heavy-atom nucleus has been demonstrated and analyzed in detail in Refs.~\cite{Skripnikov:15a,Skripnikov:17a}. Next, higher-order correlation effects were taken as the difference between values of the constants under consideration calculated within the CCSDT and CCSD(T) methods correlating 27 electrons of RaF and using the SBas basis set from Ref.~\cite{skripnikov2020nuclear}. Additionally, we calculated the contribution of even higher-order correlation effects by comparing the results obtained from CCSDT(Q) and CCSDT calculations, which correlate 27 electrons of RaF. We used the compact basis set comprising $[8s\, 8p\, 7d\, 4f]$ contracted Gaussian functions for Ra, while for F, we employed the aug-cc-pVDZ-DK basis set~\cite{Kendall1992,de2001parallel} and employed the two-component two-step approach within the generalized relativistic effective core potential (GRECP) theory~\cite{Titov06amin,Skripnikov15b,Skripnikov16a}. Note that we checked that within the compact basis set it is possible to reproduce the correction on perturbative triple excitations: the difference between the $E_{\rm eff}$ values calculated at the CCSD(T) and CCSD levels within the compact basis set, $-$0.8 GV/cm, is in good agreement with the value $-$0.9 GV/cm obtained in the much larger extAE3Z basis set. Next, for the case of $E_{\rm eff}$, $W_{P,T}$, a basis set correction, calculated at the 69e-CCSD(T) level within the Dirac-Coulomb Hamiltonian was added. Here, the extended number of basis functions in the extAE4Z basis set with respect to extAE3Z was accounted for. For the case of $W_{a}$, an equivalent correction was calculated in a similar way, but using the two-component 27e-CCSD(T) two-step approach~\cite{Titov06amin,Skripnikov15b,Skripnikov16a}.
To test the influence of further basis functions with high angular momentum for $E_{\rm eff}$, $W_{P,T}$, additional corrections were determined which capture the effect of $[15g\, 15h\, 15i\ ]$-type basis functions within the scalar-relativistic two-step approach~\cite{Titov06amin,Skripnikov15b,Skripnikov16a} and the 37e-CCSD(T) method. 
The Gaunt interelectron contribution was calculated at the self-consistent level and then rescaled by the factor 1.4 to account for correlation effects. The factor has been calculated as the ratio of the results obtained at the all-electron CCSD(T) and Dirac-Hartree-Fock levels of theory employing the large basis set used in the main calculation. To test the applicability of such a scaling approach we also calculated the Gaunt interaction contribution using a similar scheme, but within a smaller basis set of double-zeta quality \cite{dyall2016relativistic,Dyall:2009}. The obtained Gaunt contribution to $E_{\rm eff}$ of $-$0.85~GV/cm is almost equivalent to the difference of the CCSD(T) values calculated with the rigorous Dirac-Coulomb-Gaunt and Dirac-Coulomb Hamiltonians ($-$0.83~GV/cm). This therefore justifies the scaling procedure that was employed. Two-electron Gaunt interaction integrals over molecular bispinors were calculated within the code developed in Refs. \cite{Maison:20a,Maison:2019}. The calculations described above were performed at a fixed internuclear distance of 2.24~\AA, which corresponds to the equilibrium distance of the electronic ground state~\cite{kudashov2014ab,skripnikov2020nuclear}.
Finally, a vibrational correction was implemented to the considered molecular constants for the ground vibrational levels of RaF using the two-step two-component 37e-CCSD(T) approach similar to Ref.~\cite{skripnikov2020nuclear}.
In the calculations described above, a Gaussian nuclear charge distribution model~\cite{Visscher:1997} was used.

\begin{table}
\caption{Theoretical contributions to the $E_{\rm eff}$ (in units of GV/cm), $W_{P,T}$ (in units of $h$~kHz) and $W_a$ (in units of $h$~Hz) molecular parameters.}
\begin{tabular}{lrrr}
\hline
\hline
Contribution     & $E_{\rm eff}$ & $W_{P,T}$  & $W_a$ \\
\hline
CCSD(T)          & -53.9 & -145.2  & 1707 \\
CCSDT(Q) $-$ CCSD(T) & -0.1  & -0.3  & 3  \\
Basis correction       & 0.0   & -0.1 & -3  \\
Gaunt          & 0.9   & 1.4  & -16  \\
Vibr.          & -0.1  & -0.4 &  4   \\
\\
Total          & -53.3[9] & -144.3[14] &  1694[17] \\
\hline
\hline
\end{tabular}
\label{TcalculationsViolation}
\end{table}

The calculated values of $E_{\rm eff}$, $W_{P,T}$ and $W_{a}$ are given in Table~\ref{TcalculationsViolation}. High-order correlation effects given in the ``CCSDT(Q) $-$ CCSD(T)'' lines can be seen to be quite small in addition to basis set corrections. These two sources of the theoretical uncertainty are therefore almost negligible for the present case. A special note should be made concerning the Gaunt interelectron interaction contribution to $E_{\rm eff}$. The one-electron operator (Eq.~\ref{Wd2}) is only valid within the Dirac-Coulomb Hamiltonian~\cite{Lindroth89}. When the Gaunt interaction is included in the molecular Hamiltonian, the expression (Eq.~\ref{Wd2}) should be replaced by a two-electron operator. This is not trivial to realize computationally for molecules. Therefore, following a previous analysis~\cite{skripnikov2016combined}, the whole ``Gaunt'' contribution was included in the uncertainty of $E_{\rm eff}$. This contribution represents the dominant source of uncertainty. The effect of this approximate method for calculating the Gaunt contribution was also included in the theoretical uncertainty of the other calculated constants. The final uncertainty estimation of $E_{\rm eff}$, $W_{P,T}$ and $W_{a}$ values given in the main text includes: (i) contribution of higher-order correlation effects estimated as the value given in the ``CCSDT(Q) $-$ CCSD(T)'' line of Table~\ref{TcalculationsViolation}; (ii) effects of further extending the basis set which are expected to be at the level of the values given in the ``Basis correction'' line of Table~\ref{TcalculationsViolation} and (iii) effects of the Gaunt (Breit) interelectron interaction described above;
(iv) The contribution of quantum electrodynamics (QED) effects, which has been estimated as the difference in calculated values of $E_{\rm eff}$, $W_{P,T}$, and $W_{a}$ obtained using the Dirac-Coulomb Hamiltonian with and without the inclusion of the model QED operator~\cite{Shabaev:13} in the formulation in Ref.~\cite{Skripnikov2021}. It is important to note that we did not include the obtained QED contributions ($E_{\rm eff}$: 0.2 GV/cm, $W_{P,T}$: -0.2 kHz, $W_{a}$: 3 Hz) in the final values of the calculated constants, as the approach used is not a rigorous QED treatment. It can however still be used to estimate the order of magnitude of the QED effects.
The final uncertainty is calculated as the root of the sum of the squares of these uncertainties. Contributions to these uncertainties from nuclear structure effects are not considered and included due to the lack of corresponding nuclear structure calculations.

The effective Hamiltonian of the $\mathcal{P,T}$-odd interaction of the nuclear Schiff moment with electrons that contains a finite nuclear size correction is given by the following expression~\cite{GFreview,flambaum2012screening}:
\begin{equation}
     H^{{\rm eff},2}= W_S \mathbf{S''} \cdot \mathbf{n},
 \label{HXnew}
\end{equation}
where $S''$ is the corrected nuclear Schiff moment \cite{flambaum2012screening} and $W_S$ can be calculated as:
\begin{equation}
    W_S=\langle\Psi|\sum_a  \frac{3\mathbf{r_a} \cdot \mathbf{n}}{B}\rho_{N}|\Psi\rangle,
  \label{W_Sexpress}
\end{equation}
where $B=\int\rho_{N}(r)r^4dr$.
Direct use of Eq.~(\ref{W_Sexpress}) requires very large Gaussian-type basis sets. Alternatively, the relation $W_S\approx 6X/r^{sp}$ can be used, where the coefficient $r^{sp}$ was calculated analytically in Ref.~\cite{Flambaum:2019b} and the molecular parameter $X$ can be computed as follows~\cite{Hinds:80a,Sushkov:84}:
\begin{eqnarray}
X=-\frac{2\pi}{3}\langle\Psi|[\sum_i \mathbf{\nabla_i}\cdot \mathbf{n},\delta(\mathbf{R})]|\Psi\rangle.
  \label{X}
\end{eqnarray}
Following Ref.~\cite{Skripnikov:2020c}, the $X$ parameter was calculated at the 2-component CCSD(T) level using the two-step approach~\cite{Titov06amin,Skripnikov15b,Skripnikov16a}, which allowed the use of the accurate asymptotic behaviour of the wavefunction inside the nucleus. 37 electrons were included in the correlation treatment using the extAE3Z basis set. According to this calculation, correlation effects reduced the relativistic Hartree-Fock value by a factor of 1.74. The Dirac-Hartree-Fock level of theory was also used to directly calculate $W_S$ according to Eq.~(\ref{W_Sexpress}) where it was possible to use a very large basis set. The latter was constructed by modifying the extAE3Z basis set where all $s-$ and $p-$type functions were replaced by even-tempered series of Gaussian functions. Here, the Gaussian exponential parameters $\beta_i$ were calculated as $\beta_{i+1}=\beta_{i} \cdot 1.6$, where $\beta_1=1.0 \times 10^{-3}$ and the maximal $\beta_{i=64}=7.3 \times 10^9$. The final value of $W_S$ was obtained by applying the factor of 1.74 that takes into account correlation effects at the Dirac-Hartree-Fock value. The expected uncertainty of the final $W_S$ is $\sim 10 \%$, similar to that estimated in Ref.~\cite{Skripnikov:2020c} for $W_S$ constants for actinide-containing molecules. As previous works reported only the parameter $X$ to present the $W_{s}$ entries in Table 2, we computed $W_{s}$ via $W_{s} \approx 6X/r^{sp}$, using the $X$ values from those studies and $r^{sp}$ = 1.155 from Ref.~\cite{Flambaum:2019b}.

The calculated values of the symmetry-violating electronic form factors are compared with literature values in the main text. Below, we provide a brief overview of the methods used in those studies (see the corresponding references for additional details).
(i) In Ref.~\cite{kudashov2014ab}, the values of all molecular parameters of the $\mathcal{P}$-$\mathcal{P}$,$\mathcal{T}$-odd interactions were calculated for the ground electronic state of RaF using the 2-component two-step approach with the GRECP method~\cite{Titov06amin,Skripnikov15b,Skripnikov16a}. Electronic correlation effects were treated using the relativistic Fock-Space coupled cluster method with single and double excitations, along with a correction for higher-order correlation effects within the scalar-relativistic CCSD(T) approach. In the calculation~\cite{kudashov2014ab}, 19 electrons were correlated.
(ii) In Ref.~\cite{sasmal2016relativistic}, the authors calculated the values of $E_{\rm eff}$ and $W_{P,T}$ using the Dirac-Coulomb Hamiltonian and the CCSD method. They considered correlation effects for all electrons and set the virtual energy cutoff to 20 $E_h$.
(iii) In Ref.~\cite{Sunaga:2019}, the authors calculated the values of $E_{\rm eff}$ and $W_{P,T}$ using the Dirac-Coulomb Hamiltonian and the CCSD method. They set the virtual energy cutoff to 80 $E_h$. In comparison to the results of Ref.~\cite{sasmal2016relativistic}, where the so-called $\Lambda$-equations were solved to determine the values of $E_{\rm eff}$ and $W_{P,T}$ as analytical derivatives of the coupled cluster energy with respect to the added perturbation (such as the interaction of the electron EDM with the effective electric field or the scalar-pseudoscalar nucleus-electron interaction), an expectation value approach was employed in Ref.~\cite{Sunaga:2019}. There, the expectation value of a specific operator was calculated, considering only the linear terms in the CCSD wavefunction.
(iv) In Ref.~\cite{Gaul:2020}, the authors calculated the values of $E_{\rm eff}$, $W_{P,T}$, and $W_S$ using quasi-relativistic wavefunctions obtained within the zeroth-order regular approximation (ZORA). They treated electronic correlation effects using the hybrid Becke three-parameter exchange functional and the Lee, Yang, and Parr correlation functional (B3LYP)~\cite{b3lyp}. 
A related approach based on a quasi-relativistic generalized Hartree-Fock ansatz was used in Ref.~\cite{Isaev:2012} to compute $W_{a}$, the value of which was subsequently scaled to account approximately for spin-polarization effects. The corresponding scaling factor was determined from computed HFS constants as detailed in Ref.~\cite{Isaev:2012}.
(v) In Ref.~\cite{Cheng:2021} the value of $E_{\rm eff}$ was calculated using the exact 2-component atomic mean-field Hamiltonian and the CCSD(T) method. The virtual energy cutoff was set to 100 $E_h$ in the case of RaF.
(vi) In Ref.~\cite{Borschevsky:13}, the value of $W_a$ was calculated as the average of the Dirac--Hartree--Fock value and the value obtained with relativistic density functional theory, with this average then scaled by a core-polarization parameter. In density functional theory the Coulomb-attenuated B3LYP functional was employed, the parameters of which were adjusted in Ref.~\cite{Thierfelder:2010}.

The values of various nuclear effects of the $^{223,225}$Ra isotopes presented in Fig.~\ref{bw_effect}C were calculated as follows. The changes in mean-square charge radii $\delta\braket{r^2}$ were taken from~\cite{udrescu2021isotope}. The magnetic dipole moments $\mu$ and $\mu(\mathbf{r})$ are based on the measurements presented herein as well as electronic form factor calculations from Ref.~\cite{skripnikov2020nuclear}. The effect due to the electric quadrupole moment $Q_z$ was estimated using electronic form factor calculations from Ref.~\cite{petrov2020energy} and the value of the $^{223}$Ra nuclear electric quadrupole moment from Refs.~\cite{stone2005table,pyykko2008year}. The expected contributions from the anapole $\mathbf{a}$ and Schiff $\mathbf{S}$ moments were estimated using the electronic form factors calculated in this work. The value of the $^{225}$Ra anapole moment was calculated using the nuclear shell model \cite{GFreview,safronova2018search}, while for the Schiff moment, the value from Ref.~\cite{flambaum2019enhanced} was used:

\begin{eqnarray}
    S(^{225}{\rm Ra}) = 1.0 \ \bar{\theta}\mathrm{\ e \ fm^3},
\end{eqnarray}
where $\bar{\theta}$ is the $\mathcal{CP}$-violating phase of the QCD Hamiltonian. The upper limit on the Schiff moment in Fig.~\ref{bw_effect}C, is based on the limit on $\bar{\theta}$ from Ref.~\cite{graner2016reduced}. Finally, the effect of the magnetic quadrupole moment, MQM, effect in $^{223}$RaF is based on the calculations presented in Ref.~\cite{flambaum2022enhanced}.

The semi-empirical approximation 
\begin{equation}
\label{EeffHFS}
E_{\rm eff}=\alpha\sqrt{AA_d},
\end{equation}
assumes (see Ref.~\cite{Kozlov:97c} for details) that there is a proportionality relation between $E_{\rm eff}$ and a function of the HFS constants of a heavy atom-containing diatomic molecules with a $^2\Sigma_{1/2}$ electronic state. This expression allowed an estimation of $E_{\rm eff}$ for the YbF molecule~\cite{Kozlov:97c} using the experimental values of the HFS constants and a simple model of the electronic wavefunction, eliminating the need for large-scale calculations. To test this approach for RaF,  $E_{\rm eff}$ and the HFS constants were calculated at the simple Dirac-Hartree-Fock level. A value of $\alpha=0.0313$ GV/cm~MHz using Eq.~\ref{EeffHFS} was obtained. Next, Eq.~\ref{EeffHFS} was used again to extract the value of $E_{\rm eff}$ substituting the estimated value of $\alpha$ and the experimental values of the HFS constants determined here. The obtained value of $E_{\rm eff}$ in this case was found to be overestimated by about 19\% compared to the precise large-scale calculation presented here. It is however better than the pure Dirac-Hartree-Fock value of $E_{\rm eff}$, which is underestimated by 29$\%$.

There are additional similarities between the $E_{\rm eff}$ and HFS constants values. The calculated electron correlation effects contribute similarly: 29\% and 32\% for $E_{\rm eff}$ and $A_\parallel$, respectively. Correlation contributions from the innermost $1s2s2p3s3p3d$ electronic shells of Ra in RaF were computed to be 2.4\% and 2.3\% for $E_{\rm eff}$ and HFS, respectively.

\subsection{Acknowledgments}

\begin{acknowledgments}
This work was supported by the Office of Nuclear Physics, U.S. Department of Energy, under grants DE-SC0021176 and DE-SC0021179 (S.M.U., S.G.W., R.F.G.R., A.J.B.); the MISTI Global Seed Funds (S.M.U.); Deutsche Forschungsgemeinschaft (DFG, German Research Foundation) -- Projektnummer 328961117 -- SFB 1319 ELCH (A.A.B., R.B., K.G., T.G.); STFC grants ST/P004423/1 and ST/V001116/1 (M.L.B., K.T.F., H.A.P., J.R.R., J.W.); Belgian Excellence of Science (EOS) project No. 40007501 (G.N.); KU Leuven C1 project No. C14/22/104 (M.A.K., T.E.C., R.P.dG, G.N); FWO project No. G081422N (M.A.K., G.N.); International Research Infrastructures (IRI) project No. I001323N (M.A.K., T.E.C., R.P.G., A.D., S.G., L.L., G.N., B.vdB.); the European Unions Grant Agreement 654002 (ENSAR2); LISA: European Union’s H2020 Framework Programme under grant agreement no. 861198 (M.A., D.H., M.N., J.W.); The Swedish Research Council (2016-03650 and 2020-03505) (D.H., M.N.). The National Key RD Program of China (No: 2022YFA1604800) (X.F.Y.) and the National Natural Science Foundation of China (No:12027809). (X.F.Y.). Electronic structure calculations have been carried out using computing resources of the federal collective usage center Complex for Simulation and Data Processing for Mega-science Facilities at National Research Centre ``Kurchatov Institute'', http://ckp.nrcki.ru/, and partly using the computing resources of the quantum chemistry laboratory. Molecular electronic structure calculations performed at NRC ``Kurchatov Institute'' -- PNPI  have been supported by Grant No. 19-72-10019. Scalar-relativistic calculations performed at SPbU were supported by the foundation for the advancement of theoretical physics and mathematics ``BASIS'' grant according to Project No. 21-1-2-47-1.
\end{acknowledgments}

\end{document}